# Time-domain optical coherence tomography at 2 μm using GaSb-based broadband superluminescent diode

**IFTE KHAIRUL ALAM BHUIYAN,**[1, †,*] **ALEJANDRO MARTÍNEZ JIMÉNEZ,**[2] **RAMONA CERNAT,**[3] **ADRIÁN FERNÁNDEZ UCEDA,**[3] **JOONAS HILSKA,**[1] **MARKUS PEIL,**[1] **MANUEL JORGE MARQUES,**[3] **GEORGE DOBRE,**[3] **JUKKA VIHERIÄLÄ,**[4] **ADRIAN PODOLEANU,**[3] AND **MIRCEA GUINA**[1]

[1]*Optoelectronics Research Centre, Physics Unit, Tampere University, Korkeakoulunkatu 3, 33720, Tampere Finland*
[2] *Department of Physical Sciences, University de Castilla-La Mancha (UCLM), 45004 Toledo, Spain*
[3]*Applied Optics Group, Engineering, Mathematics and Physics School, University of Kent, Canterbury, CT2 7NH, UK*
[4]*Advanced Microelectronics Packaging group, Faculty of Information Technology and Communication Sciences, Tampere University, Korkeakoulunkatu 10, 33720 Tampere, Finland*
[†]*Department of Physical Sciences, Independent University, Bangladesh, Plot 16, Block B, Aftabuddin Ahmed Road, Bashundhara R/A, Dhaka – 1229, Bangladesh*

*\*ifte.alambhuiyan@tuni.fi*

**Abstract:** We report a time-domain optical coherence tomography (TD-OCT) system operating in the 2 μm spectral region, enabled by a GaSb-based superluminescent diode (SLD). The spectrum emitted by the SLD exhibits a full-width half-maximum (FWHM) of ~80 nm centred near 2.1 μm. For OCT operation, stable amplified spontaneous emission with low spectral ripple (<20%) is maintained at drive currents below 150 mA. The SLD is fiber coupled and integrated into a fiber-based Michelson interferometer. In the OCT system, the measured coherence envelope yields an axial resolution of approximately 300 μm in air and enables depth-resolved imaging of scattering paint-based coating samples. In contrast to OCT implementations at 2 μm wavelength region that commonly rely on supercontinuum sources, the use of GaSb-based SLDs offers a compact practical alternative, leveraging the maturity and scalability of electrically driven semiconductor light sources packaged in a standard "butterfly" module. This report represents the first demonstration of TD-OCT imaging at 2 μm using a GaSb-based SLD source and establishes its suitability for compact and scalable mid-IR OCT instrumentation targeting non-biological, low-water-content materials.

## 1. Introduction

Optical coherence tomography (OCT) utilizes low-coherence interferometry to perform non-invasive imaging with micrometer-scale resolution and millimeter-scale penetration depth, enabling to retrieve the 3D cross-sectional profile of tissue microstructure [1]. Axial resolution depends on the coherence length of the light source, and the penetration depth depends on the scattering and absorption of the object at the imaging wavelength. While OCT systems at 800 nm and 1060 nm are mature and have enabled significant advancement for *in vivo* medical diagnosis [2, 3], there is growing interest in imaging non-biological structures exploiting light sources emitting at longer wavelengths [4]. OCT in longer spectral region exhibits substantial advantages, as optical scattering coefficients generally can be assumed to exhibit inverse dependence on wavelength for many heterogeneous non-biological materials [5], leading to improved penetration depth compared to shorter wavelengths.

Present commercial OCT systems mostly operate at 800 nm, 1060 nm or 1300 nm, benefiting from the minimal water absorption in biological tissues and availability of mature optical components developed for telecommunication applications. While the initial development in OCT at these wavelengths have been triggered by medical applications [3, 6], there is an increasing trend to extend its application towards nondestructive testing and characterization. Application examples include damage inspection of polymer and composites [7, 8], ceramic layer characterization [9], monitoring of the thicknesses of thin coatings [10], velocity measurements of fluids [11], forensic document inspection [12], and dielectric coating inspection [13]. In these applications, the materials typically exhibit low water content, and hence they do not impose operational wavelength restrictions in terms of water absorption that is a common issue in biological samples. Additionally, while these materials under study generally have multiple scattering events resulting in noise, speckling and blurring can still compromise imaging quality at shorter wavelengths. Additionally, multiple scattering events and speckle compromise imaging quality at shorter wavelengths.

Therefore, to improve imaging fidelity, OCT operations have recently extended to the 2 µm spectral band, with already demonstrated notable outcomes [13–15]. However, these OCT systems have predominantly employed supercontinuum (SC) sources, which while providing high output power and extremely broad bandwidths, exhibit noise and bring limitation in terms of complexity, size, and cost.

As an alternative to SC sources, superluminescent diodes (SLDs) are semiconductor-based compact broadband light sources that have been utilized extensively in time domain (TD) and spectral domain (SD) OCT systems [16–18]. An SLD generates amplified spontaneous emissions (ASE) using a single- or double-pass optical channel [19]. Besides being electrically driven and benefiting from the compact footprint of a semiconductor light source, the distinctive characteristic of an SLD is the combination of a high brightness output beam that is typical for a laser diode, with a broadband spectrum emission akin to that of a light-emitting diode. An SLD has typical output-powers in the ~10–100 mW range, and their spectral features can be conveniently designed using established bandgap engineering techniques. Furthermore, when purposely designed to avoid waveguide reflections, they can emit a smooth near-Gaussian ASE spectrum with spectral modulation sufficiently suppressed within a large bandwidth (i.e. approaching 100 nm at -3dB). This leads to a clean axial point-spread function, which represents an important instrumental feature in OCT imaging. These properties together with benefits of low power consumption, availability as standard fiber-coupled modules, reduced system complexity and leveraging the scalability of semiconductor fabrication are key reasons for SLDs enabling wider spread of OCT technology.

In this work, the use of a GaSb-based SLD for TD-OCT operating in the 2 µm wavelength region is investigated. Such miniaturized light-sources have recently raised attention owing to their ability to provide broadband emission across the 2–3 µm range, supporting not only OCT but also multispecies spectroscopic applications [20–25]. In particular, we report the development of a fiber-based TD-OCT system incorporating a GaSb SLD emitting near 2.05 µm and evaluate its performance for imaging non-biological samples. The system is implemented using single-mode fiber interferometry, which simplifies optical alignment and facilitates compact integration. In our experiment, the spectral and coherence properties of the GaSb SLD are characterized and reported as a function of injection currents, aiming at quantifying their impact on the axial resolution and OCT image quality in the presence of fiber packaging penalties. Imaging experiments on layered scattering phantoms and painted samples validate the practical capability of the system. To the best of our knowledge, this represents the first demonstration of a fiber-based TD-OCT system operating at 2 µm exploiting GaSb SLDs. The results highlight the viability of GaSb SLDs as practical sources for such compact OCT systems in mid-IR region, with potential for cost-effective implementation through wafer-scale semiconductor fabrication and simplified optical architectures, opening a clear pathway toward deployment in non-destructive inspection and other real-world imaging scenarios where SC sources are impractical.

## 2. General performance assessment of 2 μm OCT systems

The axial resolution in a TD-OCT system is fundamentally determined by the coherence length of the light source, which is inversely proportional to its spectral bandwidth [3]. For a source with Gaussian spectrum, axial resolution in a medium of refractive index $n$ is approximated as :

$$\Delta z \approx \frac{2\ln(2)}{\pi} \frac{\lambda_0^2}{n \cdot \Delta\lambda} \tag{1}$$

where $\lambda_0$ is the central wavelength and $\Delta\lambda$ is the FWHM bandwidth. This estimation assumes a smooth source spectrum at the OCT interferometer input and therefore represents an upper band on the achievable axial resolution. Based on this analytical estimation, Fig. 1 illustrates the theoretical axial resolution for OCT sources centered at 2 μm as a function of spectral bandwidth for refractive indices ranging from $n$ = 1.0 (air) to higher-index media, such as, polymers, ceramics, or glasses where refractive index often ranges from ~1.3 to above ~2.5. The figure also shows an exponentially decayed trend where the axial resolution improves with increasing spectral bandwidth; the axial resolution is also better for higher refractive index media.

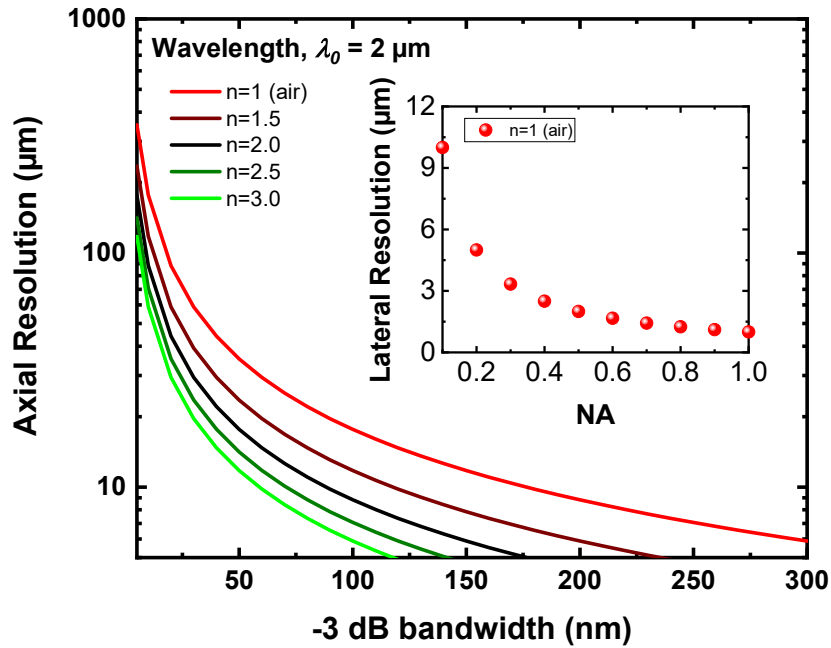

Fig. 1. Theoretical axial resolution for operation near 2 μm as a function of -3dB FWHM spectral bandwidth for different refractive indices. Inset shows lateral resolution in air (n=1) at 2 μm vs numerical aperture (NA), illustrating the lateral spot size and depth of focus.

Unlike biological tissues, where high water content dominates absorption at 2 µm and limits penetration, non-biological materials generally contain little to no water. Their optical performance is instead governed by scattering, refractive index contrast, and the presence of vibrational absorption bands. Scattering strength depends on the relative size of scattering features to wavelength, being a combination of Rayleigh scattering for sizes smaller than a wavelength (decreasing roughly with $\lambda^{-4}$ [5]) and Mie scattering for larger features than the wavelength, where scattering maybe independent to wavelength or decreases only slightly with wavelength.

Thus, in non-biological samples, both the refractive index and internal morphology determine how much backscattered light contributes to the OCT signal. Presence of multi-scattered light may also affect the measurement of the axial resolution. On the other hand, the lateral resolution of such optical system represents the minimum diameter of a focused beam depending on the optics design as well as operating wavelength, and is given by

$$r \approx \frac{\lambda_0}{2\,n\sin\theta} = \frac{\lambda_0}{2\,NA} \qquad (2)$$

where $n$ is the refractive index of the media, and $\theta$ is the half angle of the light beam that enters the objective lens. The inset of Fig. 1 shows the dependence of lateral resolution on the numerical aperture (NA) of the focusing optics for 2 µm wavelength. The lateral resolution improves with increasing NA at the expense of the depth of focus (DOF) – the axial range over which the beam remains tightly focused, so a trade-off must be considered depending on the imaging requirements. For example, doubling NA improves the lateral resolution but the depth of focus is four times reduced. At 2 µm DOF is larger than at 1.3 µm but scattering media can still limit the theoretical DOF. Overall, Fig.1 predicts that an axial resolution well below 10 µm can be obtained at 2 µm for practical refractive index ranges, provided that a source with an emission bandwidth >100 nm is available.

## 3. SLD fabrication and characterization

The SLD is based on an active region with two InGaSb/GaSb quantum wells (QWs) and has a lateral waveguide design with a J-shaped ridge waveguide geometry. The design and fabrication details have been recently reported in Alam Bhuiyan *et. al.* 2025 [23]. In current experiments a 2 mm long SLD chip was used that has 5 µm wide ridge waveguide. Along the longitudinal direction the waveguide has a straight section with a length of ~1.05 mm while the remaining waveguide exhibits an Euler bend [26] geometry with a 7° curvature terminated at the facet. For suppressing the unwanted cavity reflections at the end facet of the waveguide an $SiO_x/TiO_x$ antireflection (AR) coating was deposited with an estimated reflectivity of 0.5% anti-reflective (AR) at the curved front facet. Then, at the straight end of the waveguide a 90% high-reflective (HR) coating was deposited to enhance the ASE effect and hence the output power and the degree of coherence. At this point, it is important to note that the increase in stimulated emission comes with a decrease of the optical bandwidth and generally less smooth spectra owing to gain saturation effects. The fabricated SLD chips were mounted in a "p-side-up" configuration while the metallized n-side was bonded to the sub-mount using conductive epoxy-glue. The assembled chip was then integrated into a standard 14–pin butterfly package by Superlum Diode Ltd, with the output facet focused into a Corning SMF-28 single-mode fiber (FC/APC connector).

The SLD was driven in continuous-wave (CW) mode at room temperature (RT) and the corresponding light-current (*L-I*) and voltage-current (*V-I*) dependence are shown in Fig. 2. The output characteristics reveal the onset of ASE at a current of ~100 mA, beyond which the power increases linearly with the injection current. The maximum power measured at the fiber end was approximately 19.5 mW at a current of 1 A. For free-space operation, the same device has demonstrated a maximum power of ~40 mW at a current of 1 A, suggesting an overall fiber coupling efficiency of ~49%, a clear indication of the excellent beam quality provided by the SLD. The power saturation observed near 900–1000 mA is primarily due to saturation in gain as well as thermal rollover of the SLD device, which may be mitigated for a p-side-down mounting temperature. Furthermore, the *V–I* characteristics of the SLD measured in CW mode at room temperature reveal an accessible maximum operation voltage of ~1.3 V, which makes a strong case for simplicity of driving such devices [23, 27, 28].

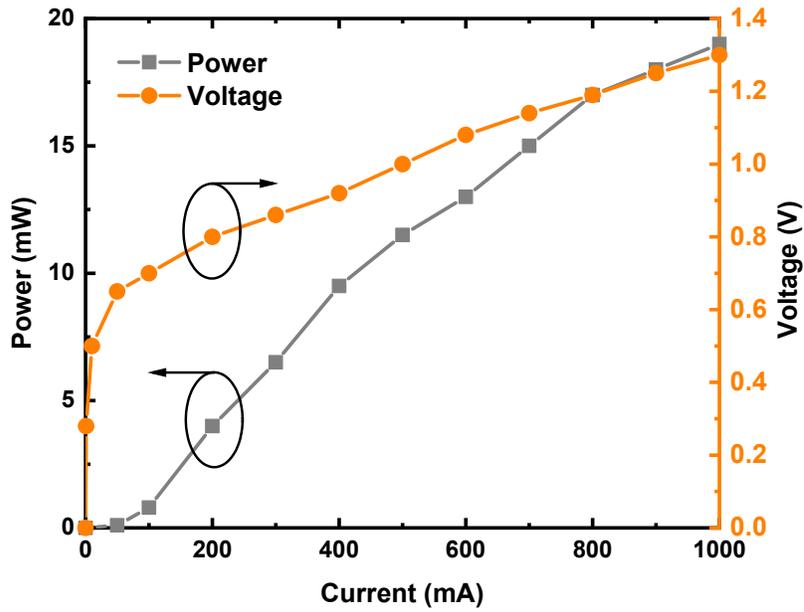

Fig. 2. (a) *L-I,* and (b) *V-I* characteristics of the pigtailed butterfly SLD module driven under CW mode in RT.

The emission spectrum was measured using Yokogawa AQ6375 optical spectrum analyzer (OSA) with a resolution of 0.1 nm, with the output coupled via an SM1950 single-mode fiber patch cord. Fig. 3 shows the measured spectrum at different currents (80–400 mA) demonstrating a broad ASE spectrum for RT operation. The inset of Fig. 3 reveals the shifts in center wavelength from ~2065 nm at lower currents (100 mA) to ~2090 nm at higher currents (400 mA), accompanied by a bandwidth (FWHM) increase from ~65 nm to ~83 nm. At the same time the output power changed from ~1 mW to ~10 mW. The measured spectral modulation (SM) is < 20% when driven up to until 150 mA, however, higher injection yields higher amplification in the double pass gain geometry resulting in increased SM of ~50% at 400 mA. Beyond this, FP cavity oscillation builds up and the device starts lasing.

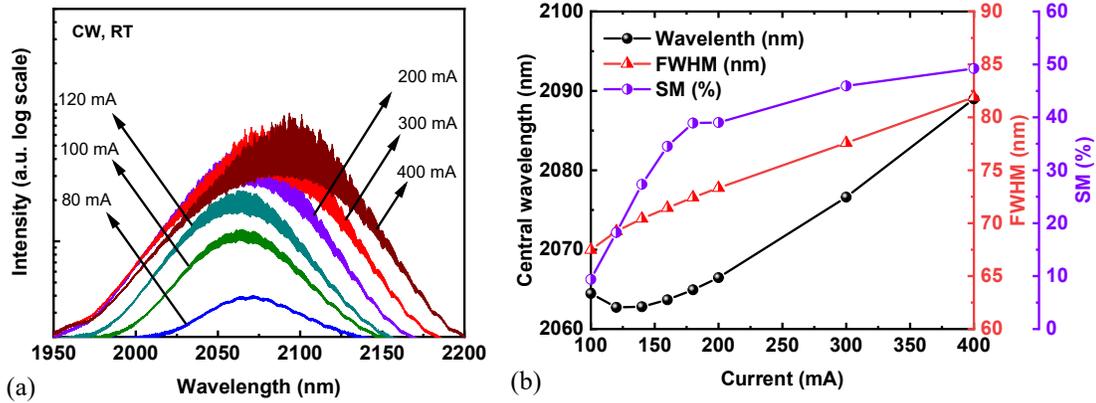

Fig. 3. ASE spectrum of the SLD measured in the free space at different injection current (0-400 mA) under CW mode in RT.

The best axial resolution in air for the broadest spectra (corresponding to a drive current of ~400 mA) was estimated to be 23 µm. The estimated axial resolution for a semi-dense media (i.e., $n$=1.5) decreases to ~19 µm and down to ~15 µm when the SLD was driven at 100 mA and 400 mA, respectively. Similarly, for more dense media (i.e., $n$=3) the resolution can be lowered to ~7.7 µm. This can provide an excellent OCT resolution, especially in semi-/non-transparent scattering materials such as polymer layers (i.e., paints), dielectric stacks, adhesives, or ceramic composites. For applications requiring even higher resolution, the SLD's emission bandwidth could be further engineered (e.g., via SLD cavity geometry) and the SM can be reduced by optimized AR/HR coatings.

## 4. TD-OCT setup and imaging

### 4.1 Interferometer setup

Fig. 4 shows the 2 µm fiber-based TD-OCT employing a Michelson interferometer configuration. The system utilizes a Thorlabs 50:50 wideband SMF-28 fiber optic coupler (TW2000R5A2B), built for broadband transmission around 2 µm (2000 ± 200 nm) and terminated with FC/APC connectors. The output of the 2 µm SLD butterfly module injects ASE into one input port of the fiber coupler (FC#1), and the signal splits into two parts: one to the reference arm and the other to the sample arm. In the sample arm, the light is first collimated using a Thorlabs's RC04APC-P01 reflective collimator (RC) to a silver reflector (450 nm - 20 µm) with a focal length of 15 mm. The XY galvo scanners, deflects the beam towards a broadband AR coated (1.65 - 3.0 µm) $CaF_2$ focusing lens L4 (LA5315-D - Ø1/2", plano-convex) with a focal length of 20 mm, which focuses the light into the sample. The galvos are positioned in the conjugate plane with the sample. In the reference arm, the 20 mm $CaF_2$ lens L1 collimates the beam, and using the same configuration the light is injected back (through L2) into the fiber. The light from reference and sample arm is recombined using another 50:50 fiber coupler (FC#2).

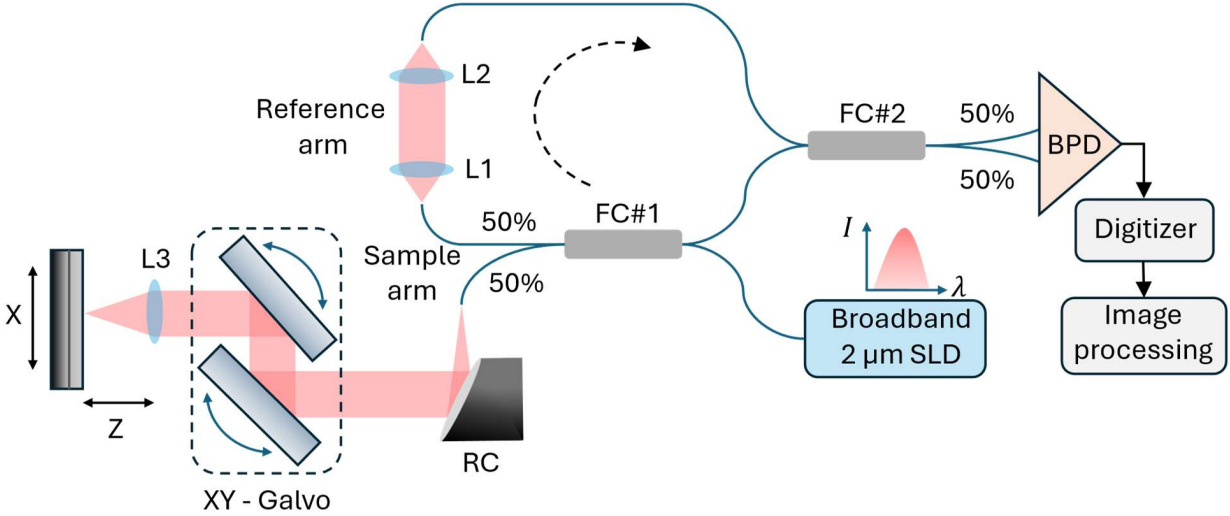

Fig. 4: Schematic of the time domain Optical Coherence Tomography (TD-OCT) system. Collimating lenses L1–L3 have identical focal lengths of 20 mm, and the parabolic reflective collimator (RC) has a focal length of 15 mm.

The balanced outputs from the FC#2 are then coupled into the custom-built balanced photodetector (BPD) using two identical Hamamastu G12183-010K extended InGaAs PIN photodiodes, with 1 A/W photosensitivity from 1.7 µm to 2.4 µm. By using a balanced detection scheme, the interferometric signal is enhanced by reducing common-mode intensity noise between reference and sample arm [29]. The signal is acquired with a 14-bit digitizer (NI PCI-6132, max 2.5 MS/s per channel) and demodulated to retrieve the fringe envelope.

*4.2 Coherence measurement*

Fig. 5 shows coherence measurement using the Michelson interferometer setup (shown in Fig. 4), when the 2 µm SLD source is driven at ~150 mA, corresponds to ~2.5 mW output power. Coherence measurement is a critical step in assessing the suitability of a light source for OCT. Unlike total output power or central wavelength, which only describe the gross emission characteristics, the coherence function directly estimates the axial resolution and image quality achievable in the TD-OCT scheme. Measuring the interferometric coherence length of the SLD therefore provides an experimental validation of its performance as a low-coherence source at 2 µm. Scanning the reference arm produces a central fringe packet around the zero-path difference position, while the sample arm contains a stable fixed mirror. The envelope of this interferogram reveals the finite coherence of the source. The measured FWHM of the envelope is approximately 0.6 mm, corresponding to an axial resolution of about 300 µm in air after accounting for the two-way propagation in the interferometer. This experimental resolution is lower than the ideal resolution predicted from the optical bandwidth alone, indicating deviations from an ideal smooth source spectrum at the interferometer input. In particular, residual cavity effects within the SLD structure and losses due to packaging optics have detrimental influence on spectral modulation and partial bandwidth reduction, which together broaden the coherence envelope and introduce secondary peaks in the axial point spread function (PSF). Despite these effects, the measurement demonstrates the essential low-coherence operation of the GaSb-based SLD at 2 µm and confirms its suitability as a practical OCT source, where the achievable axial resolution will depend primarily on spectrum shaping and fiber-coupling.

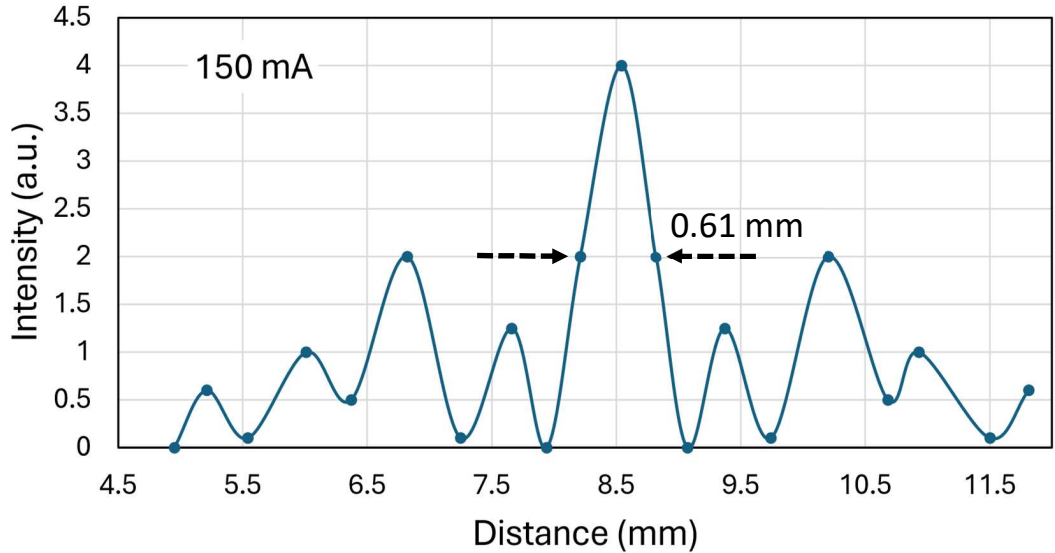

Fig. 5. Coherence measurements obtained from the interferometer setup with the SLD biased at 150 mA with the thermoelectric cooler maintaining the device at 20°C (RT).

*4.3 Setup optimization*

To optimize the scanning efficiency of the interferometer, *en-face* images of a structured test object (a British five-pence coin) were initially obtained, as shown in Fig. 6. The measurement was conducted utilizing a galvo mirror for the fast axis and a manual translation stage for the slow axis. To evaluate the performance of the configured system, confocal and OCT pictures were taken from identical targets using custom-built 100 kHz rectifier. Without rectification, both confocal and OCT images exhibit strong modulation artefacts and periodic distortions, which obscure fine details of the coin's engraved pattern. With rectification applied, these artefacts are suppressed, yielding smoother backgrounds and sharper outlines of the engraved features, confirming improved lateral fidelity of the scanning system.

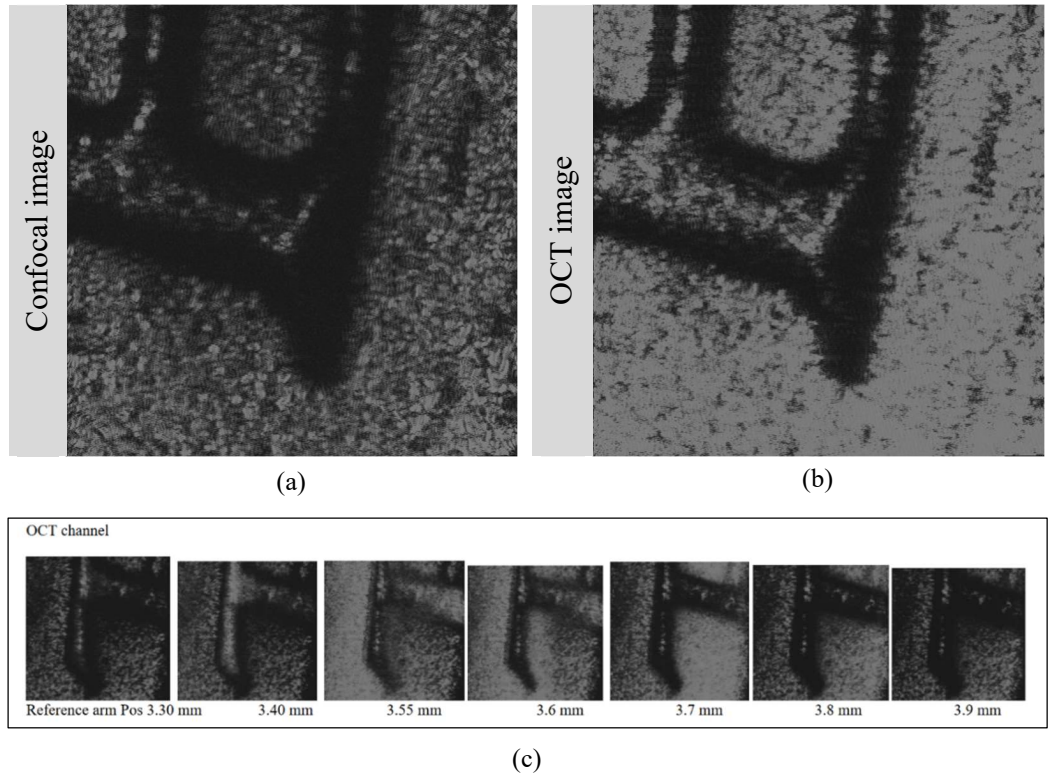

Fig. 6. *En-face* imaging of a five pence British coin by using (a) confocal and (b) OCT modalities. Each image covers an area of 2 mm x 2 mm. (c) *En-face* OCT images from OCT channels at various reference arm positions

The confocal picture in Fig. 6(a) represents an *en-face* reflection map at the objective's focal plane, where the spatial filtering of the single-mode fiber acts as a pinhole to exclude out-of-focus light. Conversely, the OCT picture in Fig. 6(b) is acquired by low-coherence interferometry, wherein the short coherence length of the GaSb SLD facilitates coherence gating: solely backscattered photons that return within the coherence length of the reference beam contribute to the interference signal. This process allows the selective detection of reflections from certain depths, generating cross-sectional pictures with micrometer-scale axial resolution. The comparison of confocal and OCT pictures confirms optical alignment and lateral resolution, while highlighting OCT's distinct advantage in revealing subsurface microstructures that conventional confocal imaging cannot resolve. Fig. 6(c) illustrates the *en-face* OCT images for various reference arm positions, showing how the system can reconstruct topography of the coin surface.

*4.4 TD-OCT imaging*

As a structured test object, we used a printed paper card (see Fig. 7) with the repeated text "UoK," onto which a thin paint coating was applied. The printed pattern provided a high-contrast lateral reference, while the paint introduced scattering and absorption representative of real-world non-biological coatings. The sample was mounted in the dedicated sample arm of the TD-OCT interferometer, positioned on a translation stage to allow controlled alignment and focusing. The probing beam from the fiber collimator was directed onto the card surface (marked with red rectangle) through the microscope objective, ensuring that both the printed features and the overlaying paint were within the optical field of view. This configuration enabled simultaneous evaluation of the system's lateral resolution by imaging the text and its axial sectioning capability by resolving subsurface structures beneath the paint. By combining a structured target with an artificial scattering layer or coatings, the setup offered a controlled yet realistic scenario to benchmark OCT performance before testing on more complex materials.

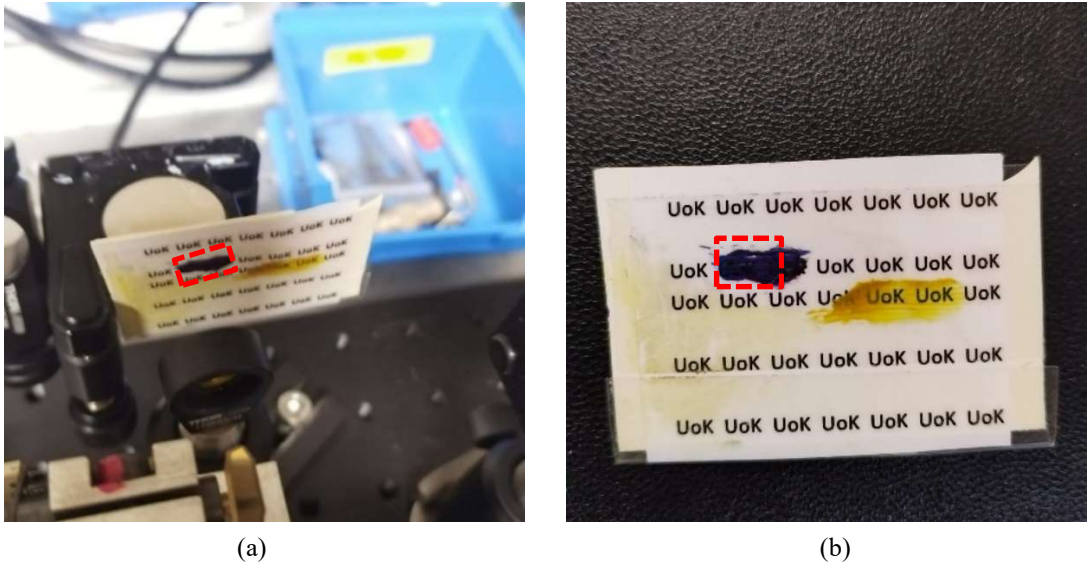

(a)  (b)

Fig. 7. Photographs of the test sample used for OCT imaging. (a) A printed card with repeated "UoK" text partially covered by a thin paint layer and positioned in the sample arm of the interferometer. (b) Front-view photograph of the same sample card. The red rectangles in both panels indicate the OCT scanning area.

The sample, as described above, is a printed card ("UoK" pattern) overlaid with a thin paint film (Edding 750 permanent marker) and imaged with a fiber-based TD-OCT using a broadband GaSb SLD centered at ~2.05 μm. *En-face* OCT slices in Fig. 8(a) and (b) were obtained by fixing reference delay at two different depths and performing lateral raster-scanning (galvo fast axis, stage slow axis). Fig. 8(a) corresponds to a plane within paint layer, while Fig. 8(b) corresponds to a plane near the printed text (ink-substrate interface). In Fig. 8(b), the buried printed text appears as bright, high-contrast features against the white paper substrate background due to local refractive-index and absorption differences at the ink–substrate interface. The mottled texture and spatial non-uniformity along the letter strokes arise from multiple scattering within the pigment coating and small-scale surface roughness introduced during paint application. The 2 μm SLD source reduces scattering relative to shorter near-infrared bands, resulting in improved penetration through the paint coating and enabling clear visualization of the buried print.

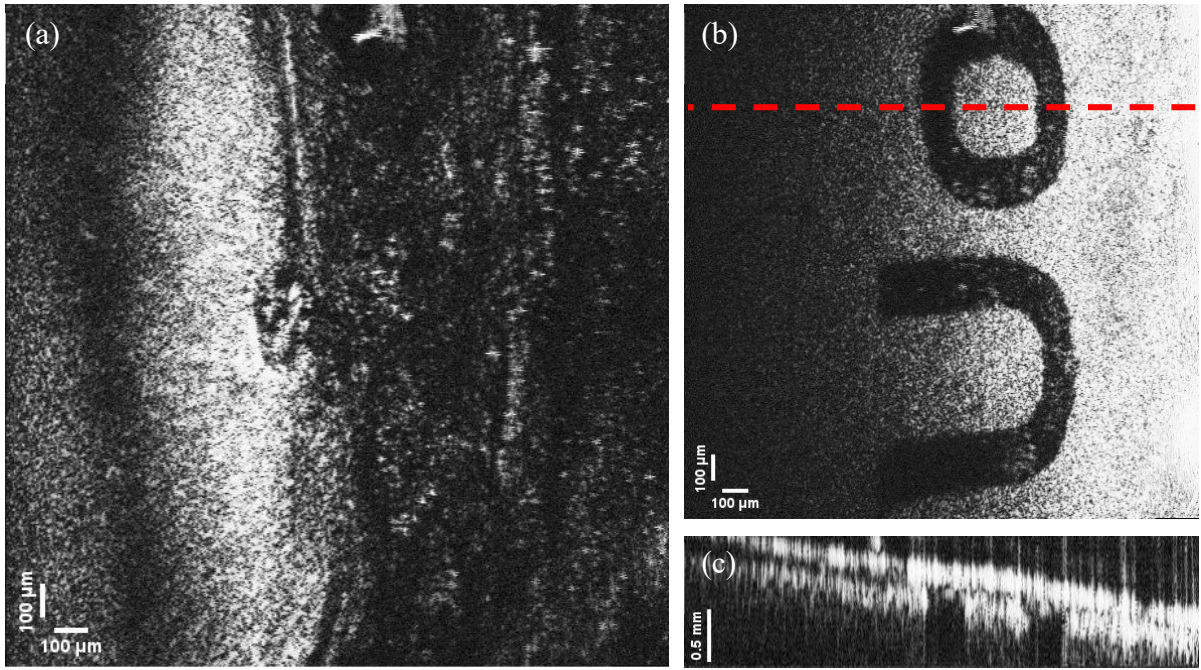

Fig. 8. Imaging of "text under paint" with the fiber-based TD-OCT using a ~2.05 µm GaSb SLD source. *En-face* slices at (a) a plane within the paint coating and (b) a plane near the buried print, revealing the "UoK" lettering as bright features beneath the paint. (c) Corresponding B-scan (along the red dashed line in (b)) reveals the air/paint surface and the paint/substrate interface; their varying separation indicates non-uniform paint thickness. Coherence gating cleanly separates surface reflections from the subsurface print.

Fig. 8 (c) is a representative B-scan acquired across the same region (see red dashed line in Fig. 8(b)) by assembling a volume of 200 *en-face* images recorded at successive depths. Axial scanning of the sample is performed using a translation stage with corresponding adjustment of the reference arm. Two dominant interfaces are evident – a strong reflection from the air/paint surfaces and a deeper band from the paint/substrate boundary. Lateral variations in the spacing between these interfaces indicate non-uniform paint thickness across the scanned area. Below the surface, signal amplitude decays with depth due to combined scattering and material absorption, while shadowing beneath the inked regions further attenuates the backscattered signals. Taken together, the *en-face* slices (see Media 1) and reconstructed B-scan demonstrate the complementary strengths of TD-OCT at 2 µm – coherence gating isolates reflections from a selected depth to reveal the buried text, while the B-scan provides a geometric readout of the layered structure (surface, paint coating, and substrate) and associated thickness variations.

## 5. Discussion

The TD-OCT experiment revealed a discrepancy between the axial resolution anticipated from the source spectrum and that recorded from the coherence envelope at drive currents under 150 mA. The bandwidth-derived estimate assumes a broad spectrum at the interferometer input, whereas the spectrum delivered by the fiber-coupled SLD module exhibits a reduced effective bandwidth together with spectral modulation. These spectral non-idealities broaden the measured coherence envelope relative to the Gaussian-limited case and introduce weak side-lobe structure, which limits the experimentally achievable axial resolution.

In TD-OCT, the axial response is usually governed by the coherence function of the source, given by the Fourier relationship between the optical spectrum and the coherence envelope [29, 30],

$$\Gamma(z) \propto \left|\int S(k)e^{i2kz}dk\right| \qquad (3)$$

The measured spectrum can be represented as a smooth envelope with a weak periodic modulation,

$$S(k) = S_0(k)[1 + m\cos(2k\Delta z + \phi)] \qquad (4)$$

where $S_0(k)$ denotes the smooth spectral envelope, $k = 2\pi/\lambda$ is the wave number, $z$ is the optical path delay, $m$ the spectral modulation depth, $\Delta z$ describes the characteristic phase delay (due to residual cavity effect in the SLD) and $\phi$ being the phase of modulation. Such spectral modulation leads to broadening of the coherence envelope and the appearance of side-lobes in the axial point-spread function [31, 32], so that axial performance depends on the spectral smoothness in addition to the spectral width. Here equations (3)–(4) are introduced to provide a qualitative interpretation of the observed coherence broadening rather than quantitative model of the measured envelope. In the present measurements, the ripple depth is approximately 20% at drive current up to 150 mA and increases to nearly 50% at 400 mA; thereby, the axial resolution does not scale directly with the increased FWHM bandwidth at higher current, despite the broader emission spectrum.

Future enhancements may be achieved through the utilization of mid-infrared-optimized single-mode fibers (such as fluoride or chalcogenide fibers) in the interferometer's fiber arms, the implementation of optimized anti-reflection coatings at the SLD facets to decrease cavity effects, the incorporation of dispersion compensating elements within the interferometer, or the application of active alignment techniques to further reduce coupling losses.

Regarding the OCT imaging, for a standard paint film with a refractive index of around 1.5–1.6, an air-calibrated resolution of ~29 μm (100 mA, 65 nm FWHM) corresponds to ~19 μm within the material, with further possibility for enhancement to ~15 μm for the wider 83 nm spectrum (400 mA). Despite the fact that these ideal values were not fully attained under the present fiber-coupled implementation at drive currents below 150 mA, the depth-resolved imaging results nonetheless demonstrate the coherence gating operation at 2 μm, where reduced scattering supports penetration through paint coatings and enables reliable distinction between surface and subsurface interfaces for thickness mapping.

## 6. Conclusion

A TD-OCT system at 2 μm regime was demonstrated using broadband ASE from a compact GaSb-based SLD. Operation at this wavelength enabled probing of a scattering and absorbing paint coatings compared to shorter-wavelength OCT. More broadly, such 2 μm OCT configuration shows promise for extending penetration depth in other low-water-content, strongly scattering materials. The broadband ASE spectrum supported axial resolutions on the order of 300 μm, sufficient to reliably resolve layered subsurface structures in scattering layer. In addition, the thermal and spectral stability of the source under CW operation ensured repeatability, an important requirement for industrial and scientific applications. With further system-level improvements, for example related to the spectral ripple reduction and use of mid-IR fibers, the GaSb SLDs can become a good candidate in the armamentarium of optical source to serve the OCT community and expand the role of OCT beyond biomedical imaging to precision metrology and materials diagnostics.

**Acknowledgement**

The authors acknowledge support from the European Commission under the Horizon 2020 program through the project Next Generation of Tunable Lasers for Optical Coherence Tomography (NETLAS, Grant Agreement No. 860807, H2020-MSCA-ITN-2019). Additional support was provided by the HORIZON EUROPE Framework Program project PhotonMed (Grant Agreement No. 101139777-2, HORIZON-KDT-JU-2023-2-RIA), Business Finland (8294/31/2023) and the Research Council of Finland Flagship Programme PREIN (Decision No. 320168). Alejandro Martinez Jimenez also acknowledges support for a Beatriz Galindo junior fellowship (BG24/00089) from the Spanish Ministry of Science, Innovation and Universities and from the University of Castilla La-Mancha. Ramona Cernat, Adrian Fernandez Uceda and Adrian Podoleanu were supported by NIHR MOORFIELDS BRC, BRC3 and in part by NIHR under Grant BRC4-05-RB413-302, at UCL Institute of Ophthalmology and Moorfields Eye Hospital. The authors wish to thank Alexander Chamorovskiy and SUPERLUM (https://superlum.ie/) for collaborating in making fiber coupled butterfly module. The authors also wish to thank Pete Verrall for making custom balanced photodetector for OCT setup.

**Disclosures Statement**

The authors declare no conflicts of interest.